\documentclass[showpacs,showkeys,preprintnumbers,eqsecnum,prd,twocolumn]{revtex4}  

\usepackage{graphicx}

\def\beq{\begin{equation}}
\def\eeq{\end{equation}}
\def\rmd{d}
\def\rmd{{\rm d}} 
\def\rmD{{\rm D}}

\begin{document}

\title{Particle dynamics and deviation effects in the field of a strong electromagnetic wave}

\author{Donato Bini}
  \affiliation{
Istituto per le Applicazioni del Calcolo ``M. Picone,'' CNR, I-00185 Rome, Italy and\\
ICRA, ``Sapienza" University of Rome, I-00185 Rome, Italy
}

\author{Andrea Geralico}
  \affiliation{Physics Department and ICRA, ``Sapienza" University of Rome, I-00185 Rome, Italy}

\author{Maria Haney}
  \affiliation{Department of Astronomy and Astrophysics, Tata Institute of Fundamental Research, Mumbai 400005, India}
  
\author{Antonello Ortolan}
  \affiliation{INFN - National Laboratories of Legnaro, I-35020 Legnaro (PD), Italy}

\begin{abstract}
Some strong field effects on test particle motion associated with the propagation of a plane electromagnetic wave in the exact theory of general relativity are investigated.
Two different profiles of the associated radiation flux are considered in comparison, corresponding to either constant or oscillating electric and magnetic fields with respect to a natural family of observers.
These are the most common situations to be experimentally explored, and have a well known counterpart in the flat spacetime limit. 
The resulting line elements are determined by a single metric function, which turns out to be expressed in terms of standard trigonometric functions in the case of a constant radiation flux, and in terms of special functions in the case of oscillating flux, leading to different features of test particle motion.
The world line deviation between both uncharged and charged particles on different spacetime trajectories due to the combined effect of gravitational and electromagnetic forces is studied.
The interaction of charged particles with the background radiation field is also discussed through a general relativistic description of the inverse Compton effect.
Motion as well as deviation effects on particles endowed with spin are studied too.
Special situations may occur in which the direction of the spin vector change during the interaction, leading to obsevables effects like spin-flip.
\end{abstract}

\pacs{04.20.Cv}

\maketitle

\section{Introduction}

In general relativity an electromagnetic wave corresponds to a curved spacetime, defined so that both the electromagnetic and the associated gravitational fields possess the same Killing symmetries.
There exists a large class of solutions satisfying this condition in terms of some arbitrary metric functions as well as electromagnetic stress energy tensor distribution.
The choice of either background metric or electromagnetic structure determines the features of the interaction of the wave with the surrounding matter.
A test particle scattered by the wave is expected to modify its own energy and momentum as a consequence of this interaction.
The analysis of electromagnetically induced gravitational effects in the exact theory thus requires some care and seems to be poorly investigated in the literature.

An exact solution of Einstein's field equations representing the gravitational field associated with an electromagnetic radiation field was discovered long ago (see, e.g., Ref. \cite{grifbook} and references therein).
The line element is simply described by trigonometric functions, whereas the associated electromagnetic field is constant, thus representing a very special physical situation.
In the present work we study the motion of both neutral and charged particles (together with geodesic and accelerated world line deviations) when the gravitational field of the electromagnetic wave background is expressed in terms of special functions (Mathieu functions), in comparison with the case of constant electromagnetic field mentioned above. This choice of the metric functions yields an associated electromagnetic field completely determined by a single harmonic wave, simply described in terms of standard trigonometric functions.
Notice that both cases of strong uniform as well as oscillating electromagnetic fields can be easily reproduced in a laboratory by using current high precision laser techniques.

Most of the general (mathematical) features of test particle motion as well as world line deviations in general pp-wave spacetimes have been extensively investigated in the literature (see, e.g., Refs. \cite{baldwin,aichelburg,balasin,vanholten,grifpod} and references therein).
Nevertheless, the analysis of simple explicit solutions may lead to a deeper understanding of the underlying physical properties. We take advantage of the simplicity of these solutions to perform analytical computations, especially in view of more complicated situations.
For instance, we study the interaction of charged particles with the radiation field by considering accelerated orbits with a further contribution to the acceleration proportional to the energy-momentum distribution of the wave. In a sense, during the scattering process the particle absorbs and re-emits radiation, resulting in a force term acting on the particle itself.
This is a second order effect of the scattering problem which could be relevant in the relativistic regime \cite{bgscat}.
A strong electromagnetic wave is indeed able to transfer enough energy to a charged particle for the particle to reach relativistic velocity after a short time. 
As a result, the photons of the radiation field will be upscattered by the relativistic particles in the associated inverse Compton process.
For instance, very high energy emission in pulsars is attributed to the inverse Compton scattering of soft stellar photons by energetic particles in the pulsar wind \cite{kirk}.
This analysis thus naturally leads to observable effects mostly associated with the interaction between plasmas and strong electromagnetic waves.

Deviations from geodesic motion can also be due to the particle's additional structure.
We study the motion of particles endowed with spin according to the Mathisson-Papapetrou-Dixon model \cite{math37,papa51,tulc59,dixon64,dixon69,dixon70,dixon73,dixon74}.
The high symmetry of the background spacetime allows to get explicit solutions for an arbitrary profile of the radiation flux.
We then discuss the shape deformation of a bunch of particles initially at rest due to their interaction with the electromagnetic wave by considering different kinds of radiation fields as well as interactions associated with the particle's additional properties.
To the best of our knowledge, such a comparative analysis has not received enough attention in the literature and represents an original contribution of the present work.

Our paper is structured as follows. In Sect. 2 we review  the solutions of Einstein-Maxwell equations representing an electromagnetic plane wave in the strong field regime. The motion of test particles (neutral, charged, spinning) is studied in Sect. 3, with a particular focus on inverse Compton scattering of charged particles. Deviations from geodesic motion are calculated explicitly in Sect. 4 for particles endowed with charge or spin. Finally, we draw our conclusions and suggest possible applications of the present analysis.

\section{The background of a strong electromagnetic plane wave}

The gravitational field associated with an electromagnetic plane wave is given by (see, e.g., Ref. \cite{szekeres,grif})
\beq
\label{radfield}
\rmd s^2 = -2\rmd u \rmd v + H^2(u)(\rmd x^2+\rmd y^2)\,, 
\eeq
written in the Rosen form, i.e., using coordinates $x^\alpha=(u,v,x,y)$.
In order to avoid coordinate singularities, we limit our considerations to the interval $u\in [0,u_B]$ where $u_B<u_*$, being $H(0)=1$ and $H(u_*)=0$.

Let the electromagnetic potential 1-form $A^\flat$ be aligned with a single spatial direction, e.g., the $x$-axis, namely 
\beq
A^\flat=h(u)\,\rmd x\,,
\eeq
so that the Faraday 2-form $F^\flat =\rmd A^\flat $ turns out to be
\beq
F=h'(u)\, \rmd u \wedge \rmd x\,,
\eeq
where a prime denotes differentiation with respect to $u$.
The associated energy-momentum tensor is then
\beq
\label{Tmunuem}
T=\Phi^2k\otimes k \,,\qquad 
\Phi=\sqrt{2}\frac{h'}{H}\,,\qquad
k=\partial_v\,,
\eeq
where $\Phi$ represents the flux of the radiation field.
Such a spacetime admits the following Killing vectors 
\begin{eqnarray}
\label{killing}
\xi_{(1)}&=&\partial_v\,,\quad
\xi_{(2)}=\partial_x\,,\quad
\xi_{(3)}=\partial_y\,,\nonumber\\
\xi_{(4)}&=&-y\partial_x+x\partial_y\,,\nonumber\\	
\xi_{(5)}&=&x\partial_v+\int^u\frac{\rmd u'}{H(u')^2}\partial_x\,,\nonumber\\	
\xi_{(6)}&=&y\partial_v+\int^u\frac{\rmd u'}{H(u')^2}\partial_y\,.
\end{eqnarray}
The trasformation of the metric (\ref{radfield}) to the more familiar Brinkmann form is shown in Appendix A.

Einstein's equations $G_{\mu\nu}=8\pi T_{\mu\nu}$ reduce to the single condition
\beq
\label{eqHh}
H'' + \frac{h'{}^2}{H}=H'' + \frac{\Phi^2}{2}H=0\,,
\eeq
for the two unknown functions $H$ and $h$.
In order to determine $H$ and $h$ uniquely, one has to provide a further relation between them.
Alternatively, one can assume that one of $H$ or $h$ is a known function of $u$.
If one treats $H$ as the known function, i.e., if one fixes the background gravitational field, Eq. (\ref{eqHh}) reduces to a first order linear differential equation for $h$, whose solution can be formally written as
\beq
\label{hsolgen}
h(u)=\int_0^u \sqrt{-H(u')H''(u')}\,\rmd u'\,.
\eeq  
For every choice of $H$ one then finds a corresponding solution for $h$, i.e., the associated electromagnetic structure. 
Eq. (\ref{hsolgen}) thus identifies a class of exact solutions of the Einstein-Maxwell field equations representing a plane electromagnetic wave. 

On the other hand, if one treats $h$ as the known function, i.e., if one fixes the background electromagnetic field, Eq. (\ref{eqHh}) is a second order differential equation for $H$, which cannot be solved in general.
We discuss below two different choices of $h$ which that are of particular interest. 

The null coordinates $(u,v)$ can be related to standard Cartesian coordinates $(t,z)$ by the transformation
\beq
u=\frac{1}{\sqrt{2}}(t-z)\,,\qquad v=\frac{1}{\sqrt{2}}(t+z)\,,
\eeq
casting the metric (\ref{radfield}) in the following quasi-Cartesian form 
\beq
\label{quasicart}
\rmd s^2 = -\rmd t^2 + H^2(t-z) \,(\rmd x^2 + \rmd y^2)+ \rmd z^2\,,
\eeq
with $x$ and $y$ as above.
Moreover, we have
\beq
\partial_t =\frac{1}{\sqrt{2}}(\partial_u+\partial_v) \,,\quad \partial_z=\frac{1}{\sqrt{2}}(-\partial_u+\partial_v) \,,
\eeq
so that the direction of propagation of the electromagnetic wave turns out to be the $z$ axis, and $k=(\partial_t+\partial_z)/\sqrt{2}$.
Notice that the two directions on the wave front, i.e., the axes $x$ and $y$, are no longer equivalent, since the electromagnetic vector potential is aligned with the $x$ direction.

A family of fiducial observers at rest with respect to the coordinates $(x,y,z)$ is characterized by the $4$-velocity vector
\beq
n=\partial_t\,.
\eeq
An orthonormal spatial triad adapted to the observers $n \equiv e_0$ is given by
\beq
e_1=\partial_z\,,\qquad
e_2=\frac{1}{H}\partial_x\,,\qquad
e_3=\frac{1}{H}\partial_y\,,
\eeq
with dual $n^\flat \equiv \omega^0=-\rmd t$ and
\beq
\label{ort_frame_forms}
\omega^1=\rmd z \,,\qquad 
\omega^2= H\, \rmd x\,,\qquad 
\omega^3=H\, \rmd y\,.
\eeq
Such a frame is also parallely propagated along $e_0$, i.e., $\nabla_{e_0}e_\alpha=0$.
It is also convenient to introduce the following notation
\beq
e_+=e_2\otimes e_2+e_3\otimes e_3\,, \quad
e_{23}=e_2\wedge e_3\,.
\eeq
The associated congruence of the observer world lines is geodesic and vorticity-free, but has a nonzero expansion
\beq
\theta(n)=\frac{\Theta(n)}{2}\,e_+
=\sqrt{2}\frac{H'}{H}e_+\,.
\eeq

The frame components of $F$ are
\beq
F=\frac{\Phi}{2} [\omega^0 \wedge \omega^2 -\omega^1 \wedge \omega^2]
=n^\flat \wedge E(n) + {}^{*_{(n)}} B(n)\,,
\eeq
where the symbol $^{*_{(u)}}$ denotes the spatial dual of a spatial tensor with respect to $u$.
The electric and magnetic fields as measured by the fiducial observers $n$ are thus given by
\beq
E(n)=-\frac{\Phi}{2}\, e_2 \,,\qquad 
B(n)=\frac{\Phi}{2}\, e_3\,.
\eeq
One easily recognizes that the electromagnetic field has a wave-like behavior, since the two electromagnetic invariants both vanish, i.e.,
\beq
E(n)^2-B(n)^2=0\,,\qquad 
E(n)\cdot B(n)=0\,.
\eeq

The nonzero frame components of the Riemann tensor are
\begin{eqnarray}
R_{0202}&=&R_{0303}=R_{1212}=R_{1313}\nonumber\\
&=&-R_{0313}=-R_{0212}\nonumber\\
&=&-\frac{H''}{2H}=\frac{\Phi^2}{4}\,,
\end{eqnarray}
so that the electric (${\mathcal E}(n)$), magnetic  (${\mathcal H}(n)$) and mixed  (${\mathcal F}(n)$) parts of the Riemann tensor (see, e.g., Ref. \cite{book} for their standard definitions) are given by
\begin{eqnarray}
{\mathcal E}(n)&=&{\mathcal F}(n)
=\frac{\Phi^2}{4}e_+\nonumber\\
&=&E(n)\otimes E(n)+B(n)\otimes B(n)\,,\nonumber\\
{\mathcal H}(n)&=& -\frac{\Phi^2}{4} e_{23}
=E(n)\wedge B(n)\,.
\end{eqnarray}
Therefore, the electric and magnetic parts of the Weyl tensor are both vanishing, i.e., 
\beq
\frac12 ({\mathcal E}(n)-{\mathcal F}(n))^{\rm (TF)}=0\,,\qquad
{\rm SYM}\,{\mathcal H}(n)=0\,,
\eeq
respectively, implying that the spacetime metric is conformally flat, and hence the associated gravitational field is algebraically special and of Petrov type O.

In the following Sections 3 and 4 we will study test particle motion as well as deviation effects associated with an electromagnetic wave background.
The interaction of particles with the radiation field is different depending on their additional properties.
Besides the well known cases of neutral and charged test particles, we will consider more complicated situations, like those associated with inverse Compton scattering of charged particles and deviation effects induced by spin on particles endowed with structure. The latter two cases, to the best of our knowledge, have not been addressed in the literature.

\subsection{Electromagnetic waves with constant profile}

Let us turn to the solutions of Eq. (\ref{eqHh}).
The case of electromagnetic waves with constant profile \cite{szekeres}, used in the literature as the simplest non-trivial solution, is recovered by setting $h(u)=\sin (bu)$, with $b$ constant, leading to constant flux $\Phi=\sqrt{2}b$ of the associated radiation field and constant electric and magnetic fields as measured by the observers $n$, i.e.,
\beq
E(n)=-\frac{b}{\sqrt{2}} e_2 \,,\qquad 
B(n)=\frac{b}{\sqrt{2}} e_3\,,
\eeq
and $H(u)=\cos (bu)$.
Note that the background quantity $b$ denotes the strength of the electromagnetic wave and has the dimensions of the inverse of a length.
It is related to the frequency of the wave by $b=\sqrt{2}\omega$.

The frame components of the electric and magnetic parts of the Riemann tensor are constant as well, namely
\beq
{\mathcal E}(n)= \frac{b^2}{2}e_+\,,\qquad
{\mathcal H}(n)= -\frac{b^2}{2} e_{23}\,.
\eeq

\subsection{Electromagnetic waves with oscillating electric and magnetic fields}

Let us now choose the unknown function $h$ such that the electric and magnetic fields are both characterized by an oscillatory behavior, i.e., 
\beq
E(n)=-A \sin (bu) e_2 \,,\qquad 
B(n)=A \sin (bu) e_3\,,
\eeq
with $A$ and $b$ constants, by requiring
\beq
\label{fluxmath}
\frac{\Phi}{2}=\frac{h'}{\sqrt{2}H}=A\sin (bu)\,.
\eeq
The electric and magnetic parts of the Riemann tensor are also oscillating
\beq
{\mathcal E}(n)= A^2\sin^2(bu) e_+\,,\quad
{\mathcal H}(n)= -A^2\sin^2(bu) e_{23}\,.
\eeq

Substituting the expression (\ref{fluxmath}) for the flux of the radiation field into Eq. (\ref{eqHh}) then gives
\beq
\label{eqHhnew}
H''+2A^2\sin^2 (bu)H=0\,,
\eeq
which represents a Mathieu's differential equation with the general solution
\beq
H=c_1{\rm MathieuC}(a, q, bu)+c_2{\rm MathieuS}(a, q, bu)\,,
\eeq
in terms of even and odd general Mathieu functions with characteristic number $a={A^2}/{b^2}$ and characteristic parameter $q =a/2$ (so that they cannot be chosen independently).
The periodicity as well as asymptotic properties of the solutions of Mathieu's equation are related to the values of the characteristic exponent $\nu$, which depends on both $a$ and $q$.
For the general properties of Mathieu functions see, e.g., Ref. \cite{abrasteg}.
A short review of main definitions and basic features of Mathieu functions is given in Appendix B. 
There are many different notations and conventions used in standard mathematical textbooks, leading to different implementations in common computational softwares, like Maple\textsuperscript{TM} \cite{maple} and Mathematica\textsuperscript{TM} \cite{mathematica}.

In this paper we adopt Maple's definition of Mathieu functions. 
We discuss strong electromagnetic plane waves described by the solution (\ref{radfield})--(\ref{Tmunuem}) with functions
\begin{eqnarray}
\label{sol}
H(u) &=& {\rm MathieuC}(a, q, bu)\,,\nonumber\\
h(u) &=& \int_0^u \sqrt{2}A\sin(b x)H(x)\,\rmd x\,,
\end{eqnarray}
$H$ satisfying Eq. (\ref{eqHhnew}) with initial conditions $H(0)=1$ and $H'(0)=0$.
For the numerical integrations we fix the values of the background parameters as $A=1=b$, as an example, so that the parameters of the Mathieu's equations are $(a,q)=(1,0.5)$. The corresponding characteristic exponent is thus given by $\nu\approx1+0.2431457i$, making the solutions non-periodic.
Periodic solutions can be of course obtained by suitably choosing $a$ in such a way that $\nu$ is real and a rational number.
For instance, setting $a\approx0.2404016$ implies $\nu\approx1/2$, leading to $4\pi$-periodic solutions.

\section{Test particle motion}

Let us study the motion of test particles with 4-velocity $U$ in the background field of a strong electromagnetic plane wave.
The equations of motion are given by
\beq
\label{eqmotogen}
m a(U)^\mu={\mathcal F}(U)^\mu\,,
\eeq
where $a(U)=\nabla_UU$ is the 4-acceleration.
As seen by the observers $n$, the $4$-velocity $U^\alpha=\rmd x^\alpha/\rmd \tau$ can then be written as
\beq
\label{Udef}
U=U^\alpha \partial_\alpha =\gamma(e_0+\nu^ae_a)\,,\quad 
\gamma=(1-\delta_{ab}\nu^a\nu^b)^{-1/2}\,,
\eeq
leading to the following relations between coordinate and frame components 
\beq
\label{Ucompts}
U^t=\gamma\,,\quad
U^x=\frac{\gamma\nu^2}{H} \,,\quad
U^y=\frac{\gamma\nu^3}{H} \,,\quad
U^z= \gamma\nu^1\,. 
\eeq
It is convenient to introduce polar coordinates in the transverse plane, i.e.,
\beq
\label{polarrel}
\nu^2=\nu_\perp\cos\chi\,,\qquad 
\nu^3=\nu_\perp\sin\chi\,,
\eeq
so that the linear velocity unit spatial vector writes as
\beq
\label{polarrel2}
\nu^ae_a=\nu_\parallel e_1+ \nu_\perp (\cos \chi e_2+\sin \chi e_3)\,,
\eeq
where the notation $\nu^1=\nu_\parallel$ has been used.

Test particle motion in the field of an electromagnetic wave with constant flux has been investigated, e.g., in Ref. \cite{bgscat}, so we refer to that work for further details.

\subsection{Timelike geodesics}

The geodesic motion is governed by the equations
\begin{eqnarray}
\label{eqgeo}
\frac{\rmd \chi}{\rmd u}&=&0\,,\nonumber\\
\frac{\rmd \nu_\perp}{\rmd u}&=&\nu_\perp\frac{H'}{H}\frac{\nu_\perp^2+\nu_\parallel-1}{1-\nu_\parallel}\,,\nonumber\\
\frac{\rmd \nu_\parallel}{\rmd u}&=&-\nu_\perp^2\frac{H'}{H}\,,
\end{eqnarray}
with
\beq
\label{eqgeo2}
\gamma(1-\nu_\parallel)=\gamma_0(1-\nu_{\parallel0})\,,
\eeq
a subscript \lq\lq0'' denoting evaluation at $u=0$.
It is useful to introduce the new variables
\beq
\label{xietadef}
\xi=\nu_\perp^2\,,\quad 
\eta=1-\nu_\parallel\,,
\eeq
satisfying the following equations
\beq
\label{eqgeo3}
\frac{\rmd \xi}{\rmd u}=2\xi\frac{H'}{H}\frac{\xi-\eta}{\eta}\,,\qquad
\frac{\rmd \eta}{\rmd u}=\xi\frac{H'}{H}\,.
\eeq
The solution of the above system is straightforward: $\chi=\chi_0$ and
\begin{eqnarray}
\label{solgeofin}
\frac{\eta}{\eta_0}&=&\frac{2H^2\eta_0}{2H^2\eta_0+(1-H^2)\xi_0}\,,\nonumber\\
\frac{\xi}{\xi_0}&=&\left(\frac{\eta}{\eta_0}\right)^2\frac1{H^2}\,.
\end{eqnarray}  
Noticeably both quantities $\eta/\eta_0$ and $\xi/\xi_0$ go to zero as $u$ approaches $u_*$ for which $H(u_*)=0$.

The solution $\eta=0=\xi$ is also an equilibrium solution for Eq. (\ref{eqgeo3}).
Furthermore, Eq. (\ref{eqgeo2}) gives 
\beq
\frac{\xi-2\eta}{\eta^2}=\frac{\xi_0-2\eta_0}{\eta_0^2}\equiv C\,,
\eeq
with $C\le-1$, which is a relation between the velocities $\nu_\parallel$ and $\nu_\perp$.
It can be cast in the following form 
\beq
\label{ellipse}
C^2\left(1-\nu_\parallel+\frac{1}{C}\right)^2-C(\nu_\perp)^2=1\,,
\eeq
representing an ellipse in the 2-space ($\nu_\parallel$, $\nu_\perp$).

\begin{figure}
\begin{center}
\includegraphics[scale=0.3]{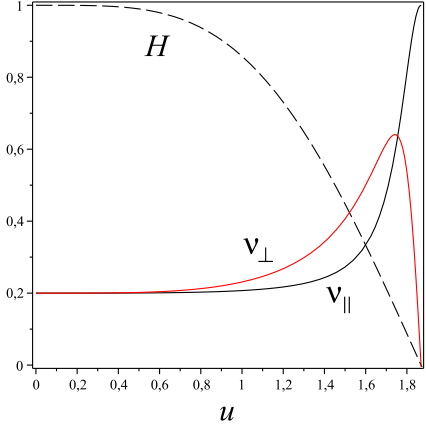}
\end{center}
\caption{The behavior of $\nu_\parallel$ and $\nu_\perp$ as functions of $u$ is shown for the geodesic case for the following choice of parameters and initial conditions:
$A=b=1$, $\nu_{\perp0}=0.2$, $\nu_{\parallel0}=0.2$.
The dashed curve represents the corresponding behavior of $H$, positively defined in the range of allowed $u$.
The features of motion in the case of constant flux are qualitatively the same.
}
\label{fig:1}
\end{figure}

The behavior of $\nu_\parallel$ and $\nu_\perp$ as functions of $u$ is shown in Fig. \ref{fig:1} for selected values of the parameters.

The corresponding parametric equations of the particle's trajectory can be obtained by further integrating the evolution equations (\ref{Ucompts}), which yield
\begin{eqnarray}
u&=&\frac{\gamma_0\eta_0}{\sqrt{2}}\tau\,,\nonumber\\
v-v_0&=&\frac1{\gamma_0^2\eta_0^2}\left(u+\gamma_0^2\xi_0\int_0^u\frac{\rmd u'}{H(u')^2}\right)\,,\nonumber\\
x-x_0&=&\sqrt{2}\frac{\xi_0^{1/2}}{\eta_0}\cos\chi_0\int_0^u\frac{\rmd u'}{H(u')^2}\,,\nonumber\\
y-y_0&=&\sqrt{2}\frac{\xi_0^{1/2}}{\eta_0}\sin\chi_0\int_0^u\frac{\rmd u'}{H(u')^2}\,.
\end{eqnarray}

\subsection{Charged particles}

Let us consider the case of an (accelerated) charged particle, with electric charge $e$.
The equations of motion (\ref{eqmotogen}) with 
\beq
\label{eqmotocharged}
{\mathcal F}(U)^\mu=eF^\mu{}_\nu U^\nu\,,
\eeq
imply
\begin{eqnarray}
\label{eqmoto}
\frac{\rmd \chi}{\rmd u}&=&\epsilon\frac{h'\sin\chi}{\gamma\nu_\perp H}\,,\nonumber\\
\frac{\rmd \nu_\perp}{\rmd u}&=&-Y\frac{\nu_\perp^2+\nu_\parallel-1}{1-\nu_\parallel}\,,\nonumber\\
\frac{\rmd \nu_\parallel}{\rmd u}&=&Y\nu_\perp\,,
\end{eqnarray}
with
\beq
Y=-\nu_\perp\frac{H'}{H}-\epsilon\frac{h'}{H}\frac{\cos\chi}{\gamma}\,,
\eeq
and $\epsilon=e/m$.
Note that Eqs. (\ref{eqgeo2}) and (\ref{ellipse}) still hold, providing a relation between the velocities $\nu_\parallel$ and $\nu_\perp$.
The above system admits the equilibrium solutions
\beq
\label{equilsol}
\nu_\parallel=1\,,\quad \nu_\perp =0 \,,\quad \chi=0,\pi\,.
\eeq
It can be analytically integrated.
In fact, equations (\ref{eqmoto}) imply
\beq
\frac{\rmd}{\rmd u}[\gamma\nu_\perp H]=-\epsilon h'\cos\chi\,,
\eeq
which together with the equation for $\chi$ gives
\beq
\gamma\nu_\perp H\sin\chi=\gamma_0\nu_{\perp0}\sin\chi_0\,,
\eeq
and
\beq
\tan\chi=\tan\chi_0\left[1-\frac{\epsilon h}{\gamma_0\nu_{\perp0}\cos\chi_0}\right]^{-1}\,.
\eeq
Substituting then into the equation for $\nu_\parallel$ yields
\begin{eqnarray}
\label{solnuparcharged}
\frac{1-\nu_\parallel}{1-\nu_{\parallel0}}&=&\left[1+\frac{(1-H^2)\nu_{\perp0}^2}{2H^2(1-\nu_{\parallel0})}\right.\nonumber\\
&&\left.
-\epsilon h\frac{2\gamma_0\nu_{\perp0}\cos\chi_0-\epsilon h}{2\gamma_0^2H^2(1-\nu_{\parallel0})}\right]^{-1}\,.
\end{eqnarray} 
Finally, the corresponding solution for $\nu_\perp$ immediately follows from Eq. (\ref{ellipse}).

\begin{figure}
\begin{center}
\includegraphics[scale=0.3]{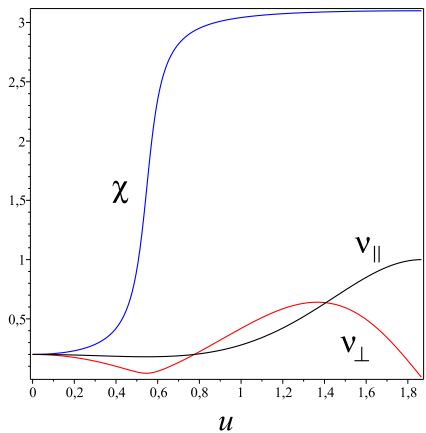}
\end{center}
\caption{The behavior of $\nu_\parallel$, $\nu_\perp$ and $\chi$ as functions of $u$ is shown for a charged particle for the same choice of parameters and initial conditions as in Fig. \ref{fig:1} with in addition $\epsilon=1$ and $\chi(0)=0.2$.
The equilibrium solution is reached at $u\approx1.865$ where $H=0$.
Recalling the polar decomposition (\ref{polarrel}) of linear velocities in the transverse plane, we then find that $\nu^2=\nu_\perp\cos\chi$ changes its sign during the evolution when $\chi$ crosses $\pi/2$, due to the electromagnetic interaction, which corresponds to the relative minimum in the curve.
}
\label{fig:2}
\end{figure}

The behavior of $\nu_\parallel$ and $\nu_\perp$ as functions of $u$ is shown in Fig. \ref{fig:2} for selected values of the parameters in the case of oscillating flux. 
The features of motion for constant flux are qualitatively the same.

\subsection{Particles undergoing inverse Compton scattering}

When a charged particle moves in a region containing an electromagnetic field, it is accelerated by the surrounding electromagnetic field itself. As a result, the charged particle radiates energy.
Part of this energy may be transferred to the photons, leading to the so called inverse Compton (IC) scattering. 
As a result, the particle feels a drag force from the emitted photons (radiation raction force).
If a particle does not radiate itself, but is embedded in an external radiation field, the IC scattering may take place as well.

Therefore, taking into account the interaction of the charged particle with the radiation field (i.e., considering effects like absorbtion and re-emission of radiation by the particle itself), the equations of motion will be modified as follows 
\beq
\label{acc_dev_eq_IC}
m a(U)=eF^\mu{}_\nu U^\nu+{\mathcal F}_{IC}(U)\,,
\eeq
where
\beq
{\mathcal F}_{IC}(U)^\alpha=-\sigma_T P(U)^\alpha{}_\beta T^{\beta}{}_{\mu} U^\mu\,, 
\eeq
is the drag force which is responsible for the IC scattering \cite{phinney,padman}.
Here $\sigma_T$ denotes the Thompson cross section of the associated process and $P(U)^{\alpha}{}_{\beta}=\delta^{\alpha}{}_{\beta}+U^{\alpha}U_{\beta}$ projects orthogonally to $U$.
It turns out that
\beq
{\mathcal F}_{IC}(U)=-\frac{\sigma_T}{2}\Phi^2\gamma^2(1-\nu_\parallel)^2\bar U\,, \qquad
\bar U\cdot \bar U=1\,,
\eeq
where 
\beq
\bar U=U-\frac{1}{\gamma(1-\nu_\parallel)}(n+e_1)
\eeq
is a unit spatial vector orthogonal to $U$, so that the magnitude of the scattering force is given by
\beq
||{\mathcal F}_{IC}(U)||=\frac{\sigma_T}{2}\Phi^2\gamma^2 (1-\nu_\parallel)^2\,.
\eeq
The equations of motion (\ref{eqmoto}) and (\ref{eqgeo2}) thus modify as
\begin{eqnarray}
\label{eqmotoPR}
\frac{\rmd \chi}{\rmd u}&=&\epsilon\frac{h'\sin\chi}{\gamma\nu_\perp H}\,,\nonumber\\
\frac{\rmd \nu_\perp}{\rmd u}&=&-Y\frac{\nu_\perp^2+\nu_\parallel-1}{1-\nu_\parallel}-\frac{\tilde\sigma_T}{2}\Phi^2\frac{\nu_\perp}{\gamma}\,,\nonumber\\
\frac{\rmd \nu_\parallel}{\rmd u}&=&Y\nu_\perp+\frac{\tilde\sigma_T}{2}\Phi^2\frac{1-\nu_\parallel}{\gamma}\,,
\end{eqnarray}
and
\beq
\label{eqmotoPR2}
\frac{\gamma(1-\nu_\parallel)}{\gamma_0(1-\nu_{\parallel0})}=\left[1+\frac{\tilde\sigma_T}{\sqrt{2}}\gamma_0(1-\nu_{\parallel0})W(u)\right]^{-1}\,,
\eeq
where $\tilde\sigma_T=\sigma_T/m$ and
\beq
\label{Wdef}
W(u)=\int_0^u\Phi(u')^2\rmd u'\,.
\eeq 
Notice that an equilibrium solution exists also in this case, still given by Eq. (\ref{equilsol}).
In the case of electromagnetic waves with constant profile we simply get
\beq
\label{intPhi1}
W(u)=2b^2u\,,
\eeq
whereas in the case of electromagnetic waves with oscillating electric and magnetic fields we find
\beq
\label{intPhi2}
W(u)=\frac{A^2}{b}[2bu-\sin(2bu)]\,.
\eeq
Furthermore, we have
\beq
\frac{\tan\chi}{\tan\chi_0}=\left[1-\epsilon\frac{1-\nu_{\parallel0}}{\nu_{\perp0}\cos\chi_0}
\int_0^u\frac{h'(u')}{\gamma(1-\nu_\parallel)}\,\rmd u'\right]^{-1}\,,
\eeq
which for a constant flux implies
\begin{eqnarray}
\frac{\tan\chi}{\tan\chi_0}&=&\left\{
1-\epsilon\frac{\sin bu}{\gamma_0\nu_{\perp0}\cos\chi_0}\left[1+\sqrt{2}b\tilde\sigma_T\gamma_0(1-\nu_{\parallel0})\right.\right.\nonumber\\
&&\left.\left.
\times\left(bu-\frac{1-\cos bu}{\sin bu}\right)
\right]
\right\}^{-1}\,.
\end{eqnarray}

The behavior of $\nu_\parallel$ and $\nu_\perp$ as functions of $u$ for fixed values of $\tilde \sigma_T$ is qualitatively the same as in Fig. \ref{fig:2}.
As the interaction strength increases, i.e., for increasing values of $\tilde \sigma_T$, the bump in the evolution of $\nu_\perp$ gets smaller and smaller, whereas $\nu_\parallel$ soon becomes relativistic. Correspondingly, the sign reversal of the frame component $\nu^2=\nu_\perp\cos\chi$ of the linear velocity in the transverse plane turns out to occur at even smaller values of $u$ (where $\chi$ crosses $\pi/2$ and $\nu_\perp$ has a relative minimum).

\subsection{Spinning particles}

The motion of massive spinning particles is decribed to first order in spin by the set of Mathisson-Papapetrou-Dixon (MPD) equations given by
\begin{eqnarray}
\label{papcoreqs1Iord}
ma(U)^\mu&\simeq&-\frac12R^{\mu}{}_{\nu\alpha\beta}U^{\nu}S^{\alpha\beta} \equiv F_{\rm (spin)}^\mu\,,\\
\label{papcoreqs2Iord}
\frac{\rmD S^{\mu\nu}}{\rmd \tau}&\simeq&0\,,
\end{eqnarray}
where $U^\alpha=\rmd x^\alpha/\rmd\tau$ denotes the timelike unit tangent vector (with proper time parameter $\tau$) to the spinning particle's ``center of mass line'' used to make a multipole reduction, and $S^{\mu\nu}$ is its antisymmetric (intrinsic angular momentum) spin tensor.
In this limit the total 4-momentum $P$ of the particle is aligned with $U$, i.e. $P^{\mu}\approx mU^\mu$, with the particle's mass $m$ remaining constant along the path.

The projection of the spin tensor into the local rest space of $U$ defines the spin vector by spatial duality
\beq
\label{spinvec}
S^\beta=\frac12\eta_\alpha{}^{\beta\gamma\delta}U^\alpha S_{\gamma\delta}=U^\alpha[{}^*S]_\alpha{}^\beta\,,
\eeq
where $\eta_{\alpha\beta\gamma\delta}=\sqrt{-g} \epsilon_{\alpha\beta\gamma\delta}$ is the unit volume 4-form and $\epsilon_{\alpha\beta\gamma\delta}$ ($\epsilon_{0123}=1$) is the Levi-Civita alternating symbol. 
The spin vector is thus parallely transported along the trajectory of the spinning particle, as in Eq. (\ref{papcoreqs2Iord}).
It is useful to introduce the signed magnitude $s$ of the spin vector
\beq
\label{sinv}
s^2=S^\beta S_\beta = \frac12 S_{\mu\nu}S^{\mu\nu}\,, 
\eeq
which is also a constant of motion.

The linearized MPD equations of motion (\ref{papcoreqs1Iord}) are formally the same as Eq. (\ref{eqmotogen}) with 
\beq
\label{eqmotospin}
{\mathcal F}(U)^\mu=F_{\rm (spin)}^\mu\,.
\eeq
A first order solution with respect to the spin can then be found in the general form
\begin{eqnarray}
\label{Uspindef}
x^\alpha&=&x_{\rm (g)}^\alpha +\tilde x^\alpha\,,\nonumber\\
U^\alpha&=&U_{\rm (g)}^\alpha +\tilde U^\alpha\,,
\end{eqnarray}
where $U_{\rm (g)}^\alpha=\rmd x_{\rm (g)}^\alpha/\rmd\tau$ denotes the unit tangent vector to a geodesic orbit, and $\tilde U^\alpha=\rmd \tilde x^\alpha/\rmd\tau$ is a deviation vector orthogonal to it (i.e., $\tilde U\cdot U_{\rm (g)}=0$, to first order in spin, as from the normalization condition $U\cdot U=-1$).
Therefore, the 4-velocity $U$ of the spinning particle has the general form (\ref{Udef}) with
\beq
\label{nuaexp}
\nu^a=\nu_{\rm (g)}^a +\tilde \nu^a\,,
\eeq 
and 
\beq
\nu_{\rm (g)}^1=\nu_{\parallel\rm (g)}\,,\quad
\nu_{\rm (g)}^2=\nu_{\perp\rm (g)}\cos\chi_0\,,\quad
\nu_{\rm (g)}^3=\nu_{\perp\rm (g)}\sin\chi_0\,.
\eeq
The spin vector must be orthogonal to $U$, so to first order we have
\beq
S=S^0n+S^1e_1+S^2e_2+S^3e_3\,,
\eeq
with
\beq
S^0=S^1\nu_{\rm (g)}^1+S^2\nu_{\rm (g)}^2+S^3\nu_{\rm (g)}^3\,.
\eeq

It is convenient to rescale the frame components of the spin vector by the particle's mass, i.e., $\sigma^a=S^a/(\sqrt{2}m)$, which have the dimensions of a length, and use the variables 
\beq
\xi_{\rm (g)}=[\nu_{\perp\rm (g)}]^2\,,\quad 
\eta_{\rm (g)}=1-\nu_{\parallel\rm (g)}\,,
\eeq
which have already been introduced in Eq. (\ref{xietadef}).
The details for the derivation of the general solution of the full set of MPD equations are given in Appendix C.
Below we simply list the main results.

The solutions for the frame components of the rescaled spin vector are 
\begin{eqnarray}
\label{solspin}
\sigma^1&=&\sigma^1_0-\left(1-\frac1H\right)\left[
\left(1-\frac1H\right)\frac{\Sigma_0}{2}\frac{\xi_0}{\eta_0}
+\sigma^1_0-\Sigma_0
\right]\,,\nonumber\\
\sigma^2&=&\sigma^2_0+\left(1-\frac1H\right)\Sigma_0\xi_0^{1/2}\cos\chi_0\,,\nonumber\\
\sigma^3&=&\sigma^3_0+\left(1-\frac1H\right)\Sigma_0\xi_0^{1/2}\sin\chi_0\,,
\end{eqnarray}
where
\beq
\Sigma_0=\sigma_0^1-\frac{\xi_0^{1/2}}{\eta_0}(\cos\chi_0\sigma_0^2+\sin\chi_0\sigma_0^3)\,,
\eeq
so that 
\beq
\frac12\left(\frac{s}{m}\right)^2=-[\sigma^1_0-\eta_0\Sigma_0]^2+[\sigma^1_0]^2+[\sigma^2_0]^2+[\sigma^3_0]^2\,.
\eeq
Their behavior is determined by the value of $\Sigma_0$, once the initial values of the geodesic velocities (i.e., $\xi_0$, $\eta_0$ and $\chi_0$) have been fixed.
In fact, let $\chi_0\in(0,\pi/2)$ and $\sigma_0^a>0$, without loss of generality.
The components $\sigma^a$ all depend on $u$ through the function $1-1/H$, which monotonically decreases from $0$ to $-\infty$ in the allowed range, for $H\to0$.
If $\Sigma_0<0$, then all $\sigma^a$ are monotonically increasing functions of $u$.
In contrast, if $\Sigma_0>0$, then all components monotonically decrease, becoming negative at a certain value of $u$.
Finally, if $\Sigma_0=0$, then $\sigma^2$ and $\sigma^3$ remain equal to their initial values, whereas $\sigma^1=\sigma_0^1/H$ monotonically increases.
Therefore, the scattering of a spinning particle by an electromagnetic wave can lead to a spin-flip effect, i.e., a sudden change of the direction of the spin vector, with its magnitude remaining constant.
This interesting feature was already discussed in the context of an interaction between gravitational waves and extended bodies with application for the observed phenomenology of glitches in pulsars \cite{quadrupgw}.

\begin{figure}
\begin{center}
\includegraphics[scale=0.3]{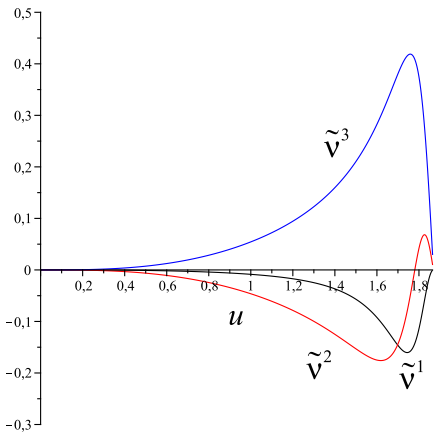}
\end{center}
\caption{The evolution of the deviation velocities $\tilde\nu^a$ during the interaction of a spinning particle with an oscillating electromagnetic field is shown for the same choice of parameters and initial conditions as in Fig. \ref{fig:1}, and with in addition $\sigma^a_0=0.1$ and $\chi_0=0.2$.
The behavior of the complete spatial velocities $\nu^a=\nu_{\rm (g)}^a +\tilde \nu^a$ given by Eq. (\ref{nuaexp}) is of course dominated by the geodesic contribution, to which the deviation velocities $\tilde\nu^a$ are only a small perturbation (the initial values of the spin components have been exaggerated to enhance the effect).
}
\label{fig:3}
\end{figure}

The solutions for the deviation velocities turn out to be 
\begin{eqnarray}
\label{soltildenu}
\tilde\nu^1&=&\frac{H'}{H}\eta_{\rm (g)}\xi_{\rm (g)}^{1/2}(\cos\chi_0\sigma_0^3-\sin\chi_0\sigma_0^2)\,,\nonumber\\
\tilde\nu^2&=&\frac{H'}{H}\eta_{\rm (g)}\left[\left(1-\frac{\xi_{\rm (g)}}{\eta_{\rm (g)}}\right)\sigma^3+\xi_{\rm (g)}^{1/2}\sin\chi_0\sigma^1\right]\,,\nonumber\\
\tilde\nu^3&=&-\frac{H'}{H}\eta_{\rm (g)}\left[\left(1-\frac{\xi_{\rm (g)}}{\eta_{\rm (g)}}\right)\sigma^2+\xi_{\rm (g)}^{1/2}\cos\chi_0\sigma^1\right]\,,\nonumber\\
\end{eqnarray}
whereas the spin-induced deviations in the transverse plane are given by
\begin{eqnarray}
\label{soltildex}
\frac{\tilde x}{\sqrt{2}}&=&\left(1-\frac{1}{H}\right)\left(\sigma^3+\Sigma\xi_{\rm (g)}^{1/2}\sin\chi_0\right)\nonumber\\
&&-(\sigma^3-\sigma_0^3)-\left(\Sigma\xi_{\rm (g)}^{1/2}-\Sigma_0\xi_0\right)\sin\chi_0\,,\nonumber\\
\frac{\tilde y}{\sqrt{2}}&=&-\left(1-\frac{1}{H}\right)\left(\sigma^2+\Sigma\xi_{\rm (g)}^{1/2}\cos\chi_0\right)\nonumber\\
&&+(\sigma^2-\sigma_0^2)+\left(\Sigma\xi_{\rm (g)}^{1/2}-\Sigma_0\xi_0\right)\cos\chi_0\,.
\end{eqnarray}
Initial conditions for the first order quantities have been chosen as $\tilde x^\alpha(0)=0=\tilde\nu^a(0)$, implying that the 4-velocity $U$ is initially tangent to the reference geodesic $U_{\rm (g)}$.
The behavior of the deviation velocities $\tilde\nu^a$ as functions of $u$, representing the corrections to the geodesic values induced by the spin, is shown in Fig. \ref{fig:3} for the case of oscillating flux.  
Note that all of deviation velocity components vanish at the end of the interaction.

The situation is somewhat different if the particle is initially at rest, i.e., $\nu_0^a\equiv0$ (or, equivalently, $\xi_0=0$ and $\eta_0=1$).
In this case the geodesic equations imply $\nu_{\rm (g)}^a\equiv0$, i.e., $\xi_{\rm (g)}\equiv0$ and $\eta_{\rm (g)}\equiv1$, for any value of $u$, so that the solutions (\ref{solspin}), (\ref{soltildenu}) and (\ref{soltildex}) to the whole set of linearized MPD equations reduce to $\sigma^a=\sigma^a_0$, 
\beq
\tilde\nu^1=0\,,\qquad
\tilde\nu^2=\sigma^3_0\frac{H'}{H}\,,\qquad
\tilde\nu^3=-\sigma^2_0\frac{H'}{H}\,,
\eeq
and
\beq
\frac{\tilde x}{\sqrt{2}}=\left(1-\frac{1}{H}\right)\sigma_0^3\,,\qquad
\frac{\tilde y}{\sqrt{2}}=-\left(1-\frac{1}{H}\right)\sigma_0^2\,,
\eeq
respectively.
By introducing the quantity
\beq
\Omega=-\frac{H'}{H}e_1\,,
\eeq
the previous solution for the deviation velocities can then be summarized by
\beq
\tilde\nu^a=[\Omega \times \sigma_0]^a\,.
\eeq
The angular velocity $\Omega$ in general diverges as $u\to u_*$, unless $\lim_{u\to u_*}(H'/H)=0$ (which is not the case for the two explicit examples considered above).
Note that both directions of $S$ and $\Omega$ are fixed, implying that the direction of $\tilde\nu$ is fixed too; its magnitude, instead, generally increases as $u\to u_*$.

\begin{figure}
\begin{center}
\includegraphics[scale=0.3]{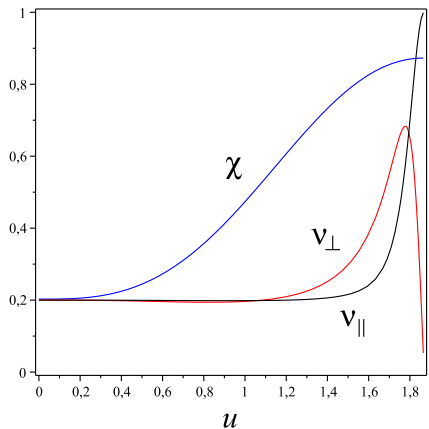}
\end{center}
\caption{The behavior of $\nu_\parallel$, $\nu_\perp$ and $\chi$ given by Eq. (\ref{polarrelspin}) as functions of $u$ is shown for a spinning particle with the same choice of parameters and initial conditions as in Fig. \ref{fig:3}.
}
\label{fig:4}
\end{figure}

Finally, in order to make a comparison with the geodesic case discussed previously, it is useful to introduce polar coordinates in the transverse plane for the linear velocities as in Eqs. (\ref{polarrel})--(\ref{polarrel2}), which  are given by
\begin{eqnarray}
\label{polarrelspin}
\nu_\parallel&=&\nu_{\parallel\rm (g)}+\tilde\nu^1\,,\nonumber\\
\nu_\perp&=&\nu_{\perp\rm (g)}+\cos\chi_0\tilde\nu^2+\sin\chi_0\tilde\nu^3\,,\nonumber\\ 
\chi&=&\chi_0+\frac{1}{\nu_{\perp\rm (g)}}(-\sin\chi_0\tilde\nu^2+\cos\chi_0\tilde\nu^3)\,,
\end{eqnarray}
to first order in spin.
Their behavior as functions of $u$ is shown in Fig. \ref{fig:4}.
The parallel component of the velocity does not significantly differ from its geodesic counterpart; the same applies to the magnitude of the transverse component, whereas its direction changes during the interaction.

\subsection{Discussion}

We have shown that the motion of a particle is modified by the dragging effects induced by its interaction with the radiation field associated with an electromagnetic wave (and the corresponding background curvature generated by the wave). 
The transverse components of the velocity orthogonal to the direction of propagation of the wave turn out to be strongly suppresses during the interaction, whereas the parallel component is enhanced, as expected. 
This is a general feature of particle motion examined here, occurring both for structureless particles and for particles endowed with additional properties (like electric charge or spin).

Consequently, if the wave propagates along the positive $z$-direction and the particle starts moving with a nonzero component of its velocity along the negative $z$-direction, there will be a certain moment during the interaction at which the dragging effects of the wave momentarily stop the particle (i.e., the parallel component of the velocity vanishes). From then on the particle and the wave both move along the same positive $z$-direction, and relativistic velocity is reached shortly thereafter. This special physical situation corresponding to a sign reversal of the parallel component of the particle's velocity can lead to measurable effects.

\begin{figure}
\begin{center}
\includegraphics[scale=0.3]{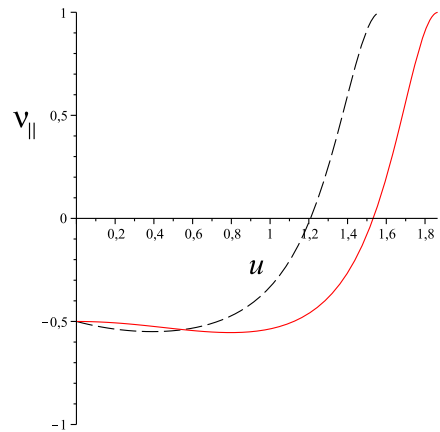}
\end{center}
\caption{The behavior of $\nu_\parallel$ as a function of $u$ is shown for a charged particle with the choice of parameters and initial conditions $\nu_{\parallel0}=-1/2$, $\nu_{\perp0}=1/2$, $\chi_0=\pi/4$, $A=1=b$ and $\epsilon=1$.
In the case of constant flux (dashed curve) the velocity changes its sign at $\bar u\approx1.2086$.
In the case of oscillating flux (solid curve), instead, the sign reversal occurs at $\bar u\approx1.5323$.
}
\label{fig:5}
\end{figure}

Consider, for instance, the case of a charged particle.
The solution for the parallel component $\nu_\parallel$ of the velocity is given by Eq. (\ref{solnuparcharged}).
By fixing, as an example, the initial data as $\nu_{\parallel0}=-1/2$, $\nu_{\perp0}=1/2$ (so that $\gamma_0=\sqrt{2}$) and $\chi_0=\pi/4$ we get
\beq
1-\nu_\parallel=\frac{11}{18}+\frac{1-2\epsilon h(1-\epsilon h)}{18H^2}\,.
\eeq
Different choices are clearly equally valid, provided that $-1<\nu_{\parallel0}<0$.
The behavior of $\nu_\parallel$ as a function of $u$ is shown in Fig. \ref{fig:5} for both constant and oscillating profiles of the radiation flux.

The condition $\nu_{\parallel}(\bar u)=0$ (i.e., backscattering of the particle) at a certain value $u=\bar u$ then gives
\beq
\label{backscatt}
H(\bar u)^2=\frac{1}{7}[1-2\epsilon h(\bar u)(1-\epsilon h(\bar u))]\,,
\eeq
which can be solved for $\bar u$, once the background solution is specified.
In the simplest case of geodesic motion (i.e., $\epsilon=0$) the above equation reduces to $H(\bar u)=1/\sqrt{7}$, which gives $b\bar u\approx1.1832$ and $b\bar u\approx1.5618$ for a constant and oscillating flux, respectively.
For charged particles and waves with constant profile, Eq. (\ref{backscatt}) implies
\beq
\label{general_sol}
\sin b\bar u=\frac{\epsilon+ \sqrt{42+13\epsilon^2}}{7+2\epsilon^2}\,.
\eeq
The above equation always admits a solutions for $\bar u$ (i.e., for every fixed value of $\epsilon$). 
For $\epsilon=0$ (neutral particles) we recover the previous result valid for the geodesic case, i.e., $\sin b\bar u=\sqrt{6/7}$, whereas in the limit $\epsilon\to \pm\infty$ we have $\sin b\bar u\to 0$.
Finally, in the case of $\epsilon \ll 1$ a first order approximation results in
\beq
\sin b\bar u\approx\sqrt{\frac{6}{7}}  + \frac{1}{7}\epsilon +O(\epsilon^2)\,,
\eeq
or, equivalently, 
\beq
b\bar u\approx 1.1832+.3780\epsilon +O(\epsilon^2)\,,
\eeq
yielding the correction to the geodesic value due to the charge.
Notice that the previous approximation does not hold for elementary particles, which have a large value of the dimensionless charge-to-mass ratio parameter $\epsilon$.
Recalling that the background parameter $b$ is simply related to the frequency of the electromagnetic wave by $b=\sqrt{2}\omega$, we expect that experiments with new generation laser devices may be conceived, in principle, to test the effect discussed above.

\section{Deviation effects}

Deviations from geodesic motion by a test particle in a given gravitational field are generally associated with the particle's additional properties, such as electric charge or spin, but also with external interacting fields, e.g., radiation fields.

\subsection{World line deviation}

Let $U$ be a reference world line and $\xi$ a generic deviation vector Lie-dragged along $U$, i.e., $[U,\xi]=0=\nabla_U \xi-\nabla_\xi U$.
By differentiating this relation along $U$ and using the index-free notation of Ref. \cite{mtw},
one gets
\begin{eqnarray}
\nabla_{UU}\xi&=&\nabla_U\nabla_\xi U
\nonumber\\
&=&[\nabla_U,\nabla_\xi]U+\nabla_\xi a(U)\nonumber\\
&=&{\mathcal R}(\ldots,U,U,\xi)+\nabla_\xi a(U)\,.
\end{eqnarray}
Denoting $\nabla_U=D/d\tau$, we have $a(U)=\nabla_UU=0$ for a geodesic orbit and hence 
\beq
\label{geo_dev_eq}
\frac{D^2\xi ^\mu}{d\tau^2}+R^{\mu}{}_{U \xi U}=0\,,
\eeq
where $X_{abc}=X_{\alpha\beta\gamma}a^\alpha b^\beta c^\gamma$.
For an orbit accelerated by a $4$-force ${\mathcal F}(U)$ we have instead $m a(U)={\mathcal F}(U)$ (see Eq. (\ref{eqmotogen})), so that the deviation equation becomes
\begin{eqnarray}
\label{acc_dev_eq_charged}
\frac{D^2\xi ^\mu}{d\tau^2}=-R^\mu{}_{U\xi U}+\frac{1}{m}\nabla_\xi {\mathcal F}(U)^\mu\,.
\end{eqnarray}

\subsubsection{Geodesic deviation}

\begin{figure}
\begin{center}
\includegraphics[scale=0.3]{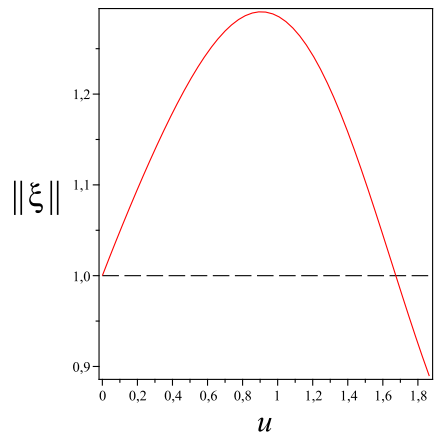}
\end{center}
\caption{The behavior of the magnitude of the deviation vector $||\xi||=\sqrt{\xi_1^2+\xi_2^2+\xi_3^2}$ is shown as a function of $u$ in the case of an oscillating flux with $\xi^1=0$, $\xi^2={\rm MathieuC}(1,0.5,u)$ and $\xi^3={\rm MathieuS}(1,0.5,u)$ (solid curve).
A bunch of particles at rest thus undergoes an oscillating shape deformation during the interaction with radiation field.
In contrast, for a constant flux ($\xi^1=0$, $\xi^2=\cos u$ and $\xi^3=\sin u$, implying that $||\xi||=1$)
the bunch preserves its shape (dashed line).
}
\label{fig:6}
\end{figure}

For the simplest scenario we consider a bunch of particles at rest with respect to the chosen coordinate system, i.e., with 4-velocity $U=e_0=\partial_t$, which is also a geodesic world line.
Let $\xi=\xi^1 e_1+\xi^2 e_2+\xi^3 e_3$ be a deviation vector from the reference world line.
Eq. (\ref{geo_dev_eq}) then gives
\beq
\label{eqsgeodev}
\frac{\rmd^2\xi_1}{\rmd\tau^2}=0\,, \qquad
\frac{\rmd^2\xi_{2,3}}{\rmd\tau^2}+\frac{\Phi^2}{4} \xi_{2,3}=0\,.
\eeq
The first equation implies that $\xi^1$ is linear in $\tau$.
However, since we are interested in deviation effects in the plane orthogonal to the direction of propagation of the electromagnetic wave, we will simply set $\xi^1=0$ hereafter.
Recalling that $\tau$ is a proper time parametrization along $\partial_t$ and hence
\beq
\rmd t = \rmd \tau=\sqrt{2} \rmd u\,,
\eeq
the second equation of (\ref{eqsgeodev}) becomes
\beq
\label{eqXdev}
\frac{\rmd^2X}{\rmd u^2}+\frac{\Phi^2}{2} X=0\,,\qquad X=\xi_{2,3}\,,
\eeq
which is equivalent to Eq. (\ref{eqHh}).
It admits the general solution
\beq
\frac{X}{H}=c_1+c_2\int^u\frac{\rmd u'}{H(u')^2}\,,
\eeq
where $c_{1,2}$ are integration constants.
Therefore, in the case of electromagnetic waves with constant profile deviations from geodesics are described by standard trigonometric functions, whereas in the case of electromagnetic waves with oscillating electric and magnetic fields deviations are governed by Mathieu functions.
As a result, in the former case the frame components of the deviation vector in the transverse plane oscillate with constant amplitude, so that a bunch of particles at rest with respect to the chosen coordinate system undergoes oscillatory deformations preserving its shape.
In the latter case, instead, the amplitude of oscillations varies with proper time, causing the bunch of particles either to spread out or to squeeze (see Fig. \ref{fig:6}).

Deviations in the transverse plane (i.e., with $\xi^1=0$) from a general timelike geodesic with $U$ given by Eq. (\ref{Udef}) and velocity components (\ref{solgeofin}) are still described by Eq. (\ref{eqXdev}), due to the symmetries of the background.

\subsubsection{Charged particle deviation}

A similar treatment for the charged case shows that the generalized deviation equation (\ref{acc_dev_eq_charged}) admits the solution  $\xi_1=0=\xi_2$, whereas $\xi_3$ still satisfies Eq. (\ref{eqXdev}). This is a consequence of the fact that the Faraday tensor has no frame components along the axis $e_3$.
Therefore, we will omit further details.

\subsubsection{Deviation of particles undergoing IC scattering}

Deviations from geodesic motion in the transverse plane due to the combined effect of both the eletromagnetic field and the IC interaction are governed by the equation
\beq
\label{eqXdev2}
\frac{\rmd^2X}{\rmd u^2}+\frac{\Phi^2}{2}\left[1-\sqrt{2}\tilde\sigma_T\gamma(1-\nu_\parallel)\frac{\rmd}{\rmd u}\ln\left(\frac{X}{H}\right)\right]X=0\,,
\eeq
where $X=\xi_{2,3}$ and we have assumed $\xi^1=0$.
Recalling the solution (\ref{eqmotoPR2}) for the quantity $\gamma(1-\nu_\parallel)$, the general solution of the above equation turns out to be
\beq
\frac{X}{H}=c_1+c_2\int^u\frac{\rmd u'}{H(u')^2}-\frac{\tilde\sigma_T}{\sqrt{2}}\gamma_0(1-\nu_{\parallel0})\int^u\frac{W(u')}{H(u')^2}\rmd u'\,,
\eeq
where higher order terms in the coupling parameter $\tilde\sigma_T$ have been neglected.
The function $W(u)$ has been defined in Eq. (\ref{Wdef}), and is given by Eq. (\ref{intPhi1}) and (\ref{intPhi2}) for both constant and oscillating flux, respectively.
In the former case we find
\beq
\int^u\frac{\rmd u'}{H(u')^2}=\frac1b\tan bu\,,
\eeq
and 
\beq
\int^u\frac{W(u')}{H(u')^2}\rmd u'=2bu\tan bu+2\ln(\cos bu)\,.
\eeq

\begin{figure}
\begin{center}
\includegraphics[scale=0.3]{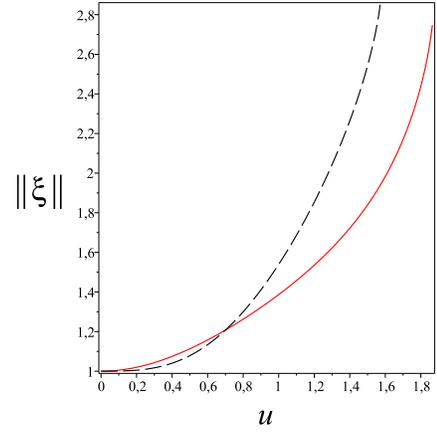}
\end{center}
\caption{The behavior of the magnitude of the deviation vector $||\xi||$ in the transverse plane (with $\xi^1=0$) is shown as a function of $u$ for the same choice of parameters as in Fig. \ref{fig:6} and $\tilde \sigma_T=1$ for both  constant flux (dashed curve) and oscillating flux (solid curve) in the presence of the IC effect.
}
\label{fig:7}
\end{figure}

Fig. \ref{fig:7} shows the behavior of the magnitude of the deviation vector in the transverse plane in both cases of constant flux and oscillating flux.
The presence of the IC effect changes the situation significantly only in the former case, with the magnitude of the deviation vector being no longer constant, but increasing with time during the interaction.

\subsection{Deviation induced by spin}

\begin{figure}
\begin{center}
\includegraphics[scale=0.3]{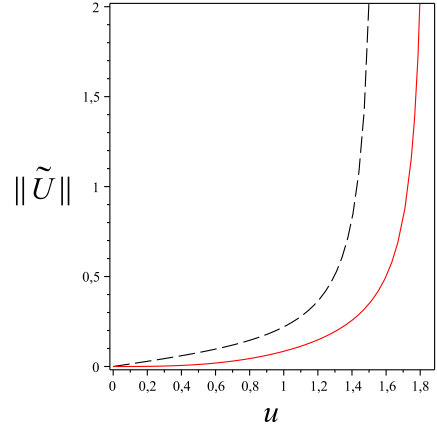}
\end{center}
\caption{The behavior of the magnitude of the deviation vector $||\tilde U||$ given by Eq. (\ref{devspinmagn}) is shown as a function of $u$ for the choice of parameters $\sigma^2_0=0.1=\sigma^3_0$ for both constant flux (dashed curve) and oscillating flux (solid curve).
}
\label{fig:8}
\end{figure}

A spinning particle deviates from geodesic motion according to Eq. (\ref{Uspindef}).
The general form of the deviation vector $\tilde U$ is given by
\begin{eqnarray}
\tilde U&=&\frac{H'}{H}\gamma_0\eta_0\left\{
\frac{1}{H}\frac{\xi_0^{1/2}}{\eta_0}(\cos\chi_0\sigma_0^3-\sin\chi_0\sigma_0^2)(n+e_1)\right.\nonumber\\
&&\left.
+\left(\sigma^3+\Sigma\xi_{\rm (g)}^{1/2}\sin\chi_0\right)e_2\right.\nonumber\\
&&\left.
-\left(\sigma^2+\Sigma\xi_{\rm (g)}^{1/2}\cos\chi_0\right)e_3
\right\}\,.\nonumber\\
\end{eqnarray}
If the particle is initially at rest, it reduces to
\beq
\tilde U=\tilde\nu^ae_a
=\frac{H'}{H}\left[\sigma^3_0e_2-\sigma^2_0e_3\right]\,,
\eeq
with magnitude
\beq
\label{devspinmagn}
||\tilde U||=\sqrt{[\sigma^2_0]^2+[\sigma^3_0]^2}\left|\frac{H'}{H}\right|\,.
\eeq
The behavior of the magnitude (\ref{devspinmagn}) of the deviation vector is shown in Fig. \ref{fig:8} for both constant and oscillating flux.

\section{Concluding remarks}

In this work we have studied deviations from geodesic motion induced by the interaction of particles endowed with additional properties (such as electric charge or spin) with the radiation field associated with the background of an exact plane electromagnetic wave.
The spacetime symmetries allow an analytical solution of the equations governing the geodesic and accelerated world line deviations for both neutral and charged test particles, as already known in the literature. 
We have then investigated more complicated situations, when the features of the scattering process are modified either by the presence of a multipolar structure of the particle (spin) or due to the inclusion of higher order effects in the interaction (inverse Compton).
We have been able to obtain explicit analytical solutions also in these cases, allowing us to discuss some interesting features which may be eventually observed, like spin-flip effects.
We have compared the two background solutions representing plane electromagnetic waves with constant profile and waves with oscillating electric and magnetic fields in the frame of a natural family of observers; these are the most common situations to be experimentally explored and have a well known counterpart in the flat spacetime limit.

The deviation equations in the transverse plane can be reduced to a single equation even if one includes higher order acceleration effects, like inverse Compton scattering.
In the case of electromagnetic waves with constant profile we have found that world line deviations of structureless particles are simply described by standard trigonometric functions with constant amplitude, whereas in the case of electromagnetic waves with oscillating electric and magnetic fields, they are governed by Mathieu functions whose amplitude varies with time.
As a result, the magnitude of the deviation vector is constant in the former case, while it increases with time in the latter. In the case of particles endowed with spin, instead, the magnitude increases during the interaction for both constant and oscillating radiation fields.
The solution of the Einstein-Maxwell equations thus highlights the parametric (non-linear) nature of the electromagnetically induced gravitational interaction.
This is a strong signature of the different gravitational content of the associated spacetime.
The study of these effects is of importance especially in view of possible experimental tests expected from future achievements of exawatt laser technologies \cite{suckewer}.

\appendix

\section{pp-wave spacetimes}

A general pp-wave spacetime associated with vacuum, Einstein-Maxwell null and pure radiation fields can be written in Brinkmann form as \cite{ES}
\beq
\label{brink}
\rmd s^2 = -2\rmd U \rmd V + K(U,X,Y)\rmd U^2+\rmd X^2+\rmd Y^2\,, 
\eeq
where $K$ is an arbitrary function of the retarded time $U$.
The corresponding spacetimes may represent strong gravitational or electromagnetic waves with arbitrary profiles.
For a pure gravitational wave the field equations give the solution
\beq
K=k_+(U)(X^2-Y^2)+2k_\times(U)XY\,,
\eeq
where $k_+(U)$ and $k_\times(U)$ are the $+$ and $\times$ polarization modes of the wave.
For a pure electromagnetic wave we have instead
\beq
K=k(U)(X^2+Y^2)\,,\qquad
k(U)\geq0\,.
\eeq

The Rosen form of the metric (\ref{radfield}) is related to the Brinkmann form (\ref{brink}) through the coordinate transformation
\begin{eqnarray}
u&=&U\,,\qquad
v=V-\frac12\frac{H'}{H}(X^2+Y^2)\,,\nonumber\\
x&=&\frac{X}{H}\,,\qquad
y=\frac{Y}{H}\,,
\end{eqnarray}
with
\beq
k=-\frac{H''}{H}=\frac{\Phi^2}{2}\,.
\eeq

\section{Mathieu functions}

We summarize below some basic properties of the Mathieu functions as well as different notations and conventions adopted in common computational softwares, like Maple\textsuperscript{TM} \cite{maple} and Mathematica\textsuperscript{TM} \cite{mathematica}.

The canonical form for Mathieu's differential equation is
\beq
\label{eqmcan}
\frac{\rmd^2y}{\rmd z^2}+(a-2q\cos2z)y=0\,,
\eeq
where the constants $a$ and $q$ are referred to as characteristic number and characteristic parameter, respectively, and are in general complex numbers.
The most general solution can be written in the form 
\beq
\label{eqmsolgen}
y(z)=c_1y_1(z)+c_2y_2(z)\,,
\eeq
where $y_{1,2}$ are two independent solutions and $c_{1,2}$ arbitrary complex constants.

According to Floquet's theorem, there exists a complex valued solution of Eq. (\ref{eqmcan}) of the form
\beq
\label{floquetsol}
y_1(z)=e^{i\nu z}p(z)\,,
\eeq
where $\nu=\nu(a,q)$ is in general a complex number called characteristic exponent, and $p(z)$ is a periodic function of $z$ with period $\pi$.
The above solution is bounded for $z\to\infty$, unless $\nu$ is a complex number, for which it is unbounded.
Its periodicity depends on the value of $\nu$: $y_1(z)$ is non-periodic if $\nu$ is complex or even real but not a rational number; if $\nu$ is a rational number, i.e., $\nu=m/n$, then $y_1(z)$ is periodic of period at most $2\pi n$; finally, if $\nu$ is a real integer, $y_1(z)$ is a periodic function with period $\pi$ or $2\pi$.
Periodic solutions of Mathieu's equation are called Mathieu functions of the first kind, or, more simply, Mathieu functions.
In this case, $a$ and $q$ cannot be given independently, because periodicity requires that they fulfill the equation $\nu(a,q)=n$, with $n$ integer. All values of $a$ satisfying the latter condition for fixed values of $q$ are called characteristic values.
It turns out that $y_1(z)$ is periodic with period $\pi$ for even $n$ and periodic with period $2\pi$ for odd $n$.
In most textbooks periodic solutions which are even functions of $z$ are denoted by $ce_n(z,q)$ (cosine elliptic), whereas $se_n(z,q)$ (sine elliptic) are odd functions. Furthermore, $ce_{2n}(z,q)$ and $se_{2n+1}(z,q)$ have period $\pi$, while $ce_{2n+1}(z,q)$ and $se_{2n+2}(z,q)$ have period $2\pi$. However, different normalization conventions are adopted in the literature.
On the other hand, if the parameters $a$ and $q$ are fixed independently (e.g., in the case of the parametric oscillator), the general solution $y_1(z)$ may be periodic or not, bounded or not depending on the corresponding values of $\nu$, as discussed before.

A second independent solution of Mathieu's equation is given by $y_2(z)=y_1(-z)$, provided that the characteristic exponent is not a real integer, otherwise it is of the form $y_2(z)=czy_1(z)+f(z)$ (Mathieu functions of the second kind), where $c$ is a constant and $f(z)$ has the same periodicity properties as $y_1(z)$.
The second linearly independent solutions (necessarily not periodic) associated with cosine elliptic and sine elliptic functions are denoted by $fe_{2n+1}(z,q)$ and $ge_{2n+1}(z,q)$, respectively.

\subsection{Maple\textsuperscript{TM}}

In Maple\textsuperscript{TM} the general solution (\ref{eqmsolgen}) of Mathieu's equation (\ref{eqmcan}) is given by
\beq
y(z)=c_1{\rm MathieuC}(a,q,z)+c_2{\rm MathieuS}(a,q,z)\,,
\eeq
in terms of even and odd functions defined by
\begin{eqnarray}
{\rm MathieuC}(a,q,z)&=&\frac12\frac{y_1(z)+y_1(-z)}{y_1(0)}\,, \nonumber\\
{\rm MathieuS}(a,q,z)&=&\frac12\frac{y_1(z)-y_1(-z)}{{y'}_1(0)}\,, 
\end{eqnarray}
respectively, where the Floquet solution (\ref{floquetsol}) is denoted by $y_1(z)={\rm MathieuFloquet}(a,q,z)$.
The Mathieu cosine and Mathieu sine functions are real valued and normalized so that 
\beq
{\rm MathieuC}(a,q,0)=1\,,\quad
{\rm MathieuC}'(a,q,0)=0\,,
\eeq
and
\beq
{\rm MathieuS}(a,q,0)=0\,,\quad
{\rm MathieuS}'(a,q,0)=1\,,
\eeq
respectively, like standard trigonometric functions.
They are in general aperiodic. A noteworthy special case is $q=0$, for which
\begin{eqnarray}
{\rm MathieuC}(a,0,z)&=&\cos\sqrt{a}z\,, \nonumber\\
{\rm MathieuS}(a,0,z)&=&\frac{\sin\sqrt{a}z}{\sqrt{a}}\,.
\end{eqnarray}
For a given pair $(a,q)$, the characteristic exponent $\nu$ entering the Floquet solution (\ref{floquetsol}) is computed using the auxiliary function ${\rm MathieuExponent}(a,q)$.
For countably many values of $a$ (as a function of $q$), the Mathieu cosine and sine functions are periodic.
The corresponding characteristic values are computed by using ${\rm MathieuA}(n,q)$ and ${\rm MathieuB}(n,q)$, respectively. 

The periodic solutions of the Mathieu's equation are denoted by ${\rm MathieuCE}(n,q,z)$ and ${\rm MathieuSE}(n,q,z)$, with $n$ a non-negative integer, which are also special cases of the Mathieu cosine and sine functions, respectively.
They are defined as

\begin{widetext}

\begin{eqnarray}
\frac{{\rm MathieuCE}(n,q,z)}{{\rm MathieuCE}(n,q,0)}&=&{\rm MathieuC}\left({\rm MathieuA}(n,q),q,z\right)\,, \qquad
n=0,1,\ldots\,, \nonumber\\
\frac{{\rm MathieuSE}(n,q,z)}{{\rm MathieuSE}'(n,q,0)}&=&{\rm MathieuS}\left({\rm MathieuB}(n,q),q,z\right)\,, \qquad
n=1,2,\ldots\,.
\end{eqnarray}

\end{widetext}
If the index $n$ is even, then both MathieuCE and MathieuSE are $\pi$-periodic, otherwise they are $2\pi$-periodic. 
They assume the following special values for $q=0$
\begin{eqnarray}
{\rm MathieuCE}(n,0,z)&=&\cos nz\,, \nonumber\\
{\rm MathieuSE}(n,0,z)&=&\sin nz\,.
\end{eqnarray}

\subsection{Mathematica\textsuperscript{TM}}

Compared to Maple\textsuperscript{TM} the Mathieu cosine and Mathieu sine functions are normalized differently and are defined as
\begin{eqnarray}
{\rm MathieuC}[a,q,z]&=&\frac{y_1(z)+y_1(-z)}{2}\,, \nonumber\\
{\rm MathieuS}[a,q,z]&=&\frac{y_1(z)-y_1(-z)}{2i}\,. 
\end{eqnarray}
For instance, for $q=0$ they reduce to
\begin{eqnarray}
{\rm MathieuC}[a,0,z]&=&\cos\sqrt{a}z\,, \nonumber\\
{\rm MathieuS}[a,0,z]&=&\sin\sqrt{a}z\,.
\end{eqnarray}

\section{Solving the MPD equations for a spinning particle}

The whole set of linearized MPD equations (\ref{papcoreqs1Iord})--(\ref{papcoreqs2Iord}) for a spinning particles can be analytically solved due to the spacetime symmetries.

The transport equations for the rescaled spin vector with components $\sigma^a=S^a/(\sqrt{2}m)$ are 
\begin{eqnarray}
\label{eqsspin}
\frac{\rmd \sigma^1}{\rmd u}&=&\frac{H'}{H}(\Sigma-\sigma^1)\,,\nonumber\\
\frac{\rmd \sigma^2}{\rmd u}&=&\frac{H'}{H}\Sigma\xi_{\rm (g)}^{1/2}\cos\chi_0\,,\nonumber\\
\frac{\rmd \sigma^3}{\rmd u}&=&\frac{H'}{H}\Sigma\xi_{\rm (g)}^{1/2}\sin\chi_0\,,
\end{eqnarray}
where
\beq
\Sigma=\sigma^1-\frac{1}{H}\frac{\xi_0^{1/2}}{\eta_0}(\cos\chi_0\sigma^2+\sin\chi_0\sigma^3)\,,
\eeq
and the variables $(\xi_{\rm (g)},\eta_{\rm (g)})$ have already been introduced in Eq. (\ref{xietadef}).
The previous equations imply
\beq
\cos\chi_0\frac{\rmd \sigma^3}{\rmd u}-\sin\chi_0\frac{\rmd \sigma^2}{\rmd u}=0\,,
\eeq
whence
\beq
\sigma^3=\sigma^3_0+\tan\chi_0(\sigma^2-\sigma^2_0)\,.
\eeq
Substituting then into the remaining equations gives the final solution (\ref{solspin}).

The equations of motion (\ref{papcoreqs1Iord}) then imply the set of equations

\begin{widetext}

\begin{eqnarray}
\label{eqmspin}
\frac{\rmd \tilde\nu^1}{\rmd u}&=&-2\frac{H'}{H}\xi_{\rm (g)}^{1/2}(\cos\chi_0\tilde\nu^2+\sin\chi_0\tilde\nu^3)
+\frac{\Phi^2}{2}\eta_{\rm (g)}\xi_{\rm (g)}^{1/2}(\sin\chi_0\sigma^2-\cos\chi_0\sigma^3)\,,\nonumber\\
\frac{\rmd \tilde\nu^2}{\rmd u}&=&\frac{H'}{H}\left\{
\frac{\xi_{\rm (g)}^{3/2}}{\eta_{\rm (g)}^2}\cos\chi_0\tilde\nu^1
-\left[1-\frac{\xi_{\rm (g)}}{\eta_{\rm (g)}}(2+\cos2\chi_0)\right]\tilde\nu^2
+\frac{\xi_{\rm (g)}}{\eta_{\rm (g)}}\sin2\chi_0\tilde\nu^3
\right\}\nonumber\\
&&
-\frac{\Phi^2}{2}\eta_{\rm (g)}\left[\xi_{\rm (g)}^{1/2}\sin\chi_0\sigma^1+\left(1-\frac{\xi_{\rm (g)}}{\eta_{\rm (g)}}\right)\sigma^3\right]\,,\nonumber\\
\frac{\rmd \tilde\nu^3}{\rmd u}&=&\frac{H'}{H}\left\{
\frac{\xi_{\rm (g)}^{3/2}}{\eta_{\rm (g)}^2}\sin\chi_0\tilde\nu^1
+\frac{\xi_{\rm (g)}}{\eta_{\rm (g)}}\sin2\chi_0\tilde\nu^2
-\left[1-\frac{\xi_{\rm (g)}}{\eta_{\rm (g)}}(2-\cos2\chi_0)\right]\tilde\nu^3
\right\}\nonumber\\
&&
+\frac{\Phi^2}{2}\eta_{\rm (g)}\left[\xi_{\rm (g)}^{1/2}\cos\chi_0\sigma^1+\left(1-\frac{\xi_{\rm (g)}}{\eta_{\rm (g)}}\right)\sigma^2\right]\,,
\end{eqnarray}
whose solution proves challenging.
Fortunately, we can take advantage of the high degree of symmetry of the background spacetime, which admits the six Killing vectors (\ref{killing}) with associated conserved quantities
\beq
\label{conserveddef}
C_{(A)}=-\xi^{(A)}_{\alpha} P^\alpha +\frac12 S^{\alpha\beta}\xi^{(A)}_{\alpha;\beta}\,,\qquad
A=1,\ldots,6\,.
\eeq
We find
\begin{eqnarray}
\label{conserved}
{\mathcal C}_{(1)}&=&-\frac{\gamma_0\eta_0}{\sqrt{2}}\left\{
1-\gamma_{\rm (g)}^2\left[\left(1-\frac{\xi_{\rm (g)}}{\eta_{\rm (g)}}\right)\tilde\nu^1
-\xi_{\rm (g)}^{1/2}(\cos\chi_0\tilde\nu^2+\sin\chi_0\tilde\nu^3)\right]
\right\}\,,\nonumber\\
{\mathcal C}_{(2)}&=&\gamma_{\rm (g)}H\xi_{\rm (g)}^{1/2}\cos\chi_0\left\{
1+\gamma_{\rm (g)}^2\left[(1-\eta_{\rm (g)})\tilde\nu^1
-\frac{\xi_{\rm (g)}\sin^2\chi_0+(1-\eta_{\rm (g)})^2-1}{\xi_{\rm (g)}^{1/2}\cos\chi_0}\tilde\nu^2
+\xi_{\rm (g)}^{1/2}\sin\chi_0\tilde\nu^3\right]\right.\nonumber\\
&&\left.
-\frac{H'}{H}\frac{\eta_{\rm (g)}}{\xi_{\rm (g)}^{1/2}\cos\chi_0}\left(\sigma^3+\Sigma\xi_{\rm (g)}^{1/2}\sin\chi_0\right)
\right\}\,,\nonumber\\
{\mathcal C}_{(3)}&=&\gamma_{\rm (g)}H\xi_{\rm (g)}^{1/2}\sin\chi_0\left\{
1+\gamma_{\rm (g)}^2\left[(1-\eta_{\rm (g)})\tilde\nu^1
+\xi_{\rm (g)}^{1/2}\cos\chi_0\tilde\nu^2
-\frac{\xi_{\rm (g)}\cos^2\chi_0+(1-\eta_{\rm (g)})^2-1}{\xi_{\rm (g)}^{1/2}\sin\chi_0}\tilde\nu^3\right]\right.\nonumber\\
&&\left.
+\frac{H'}{H}\frac{\eta_{\rm (g)}}{\xi_{\rm (g)}^{1/2}\sin\chi_0}\left(\sigma^2+\Sigma\xi_{\rm (g)}^{1/2}\cos\chi_0\right)
\right\}\,,\nonumber\\
{\mathcal C}_{(4)}&=&x_{\rm (g)}{\mathcal C}_{(3)}-y_{\rm (g)}{\mathcal C}_{(2)}+\sqrt{2}\gamma_{\rm (g)}\left\{
\eta_{\rm (g)}(2-\eta_{\rm (g)})\Sigma+\xi_{\rm (g)}^{1/2}\left[
\left(\sigma^2-\frac{H\tilde y}{\sqrt{2}}\right)\cos\chi_0+\left(\sigma^3+\frac{H\tilde x}{\sqrt{2}}\right)\sin\chi_0
\right]\right\}\,,\nonumber\\
{\mathcal C}_{(5)}&=&x_0{\mathcal C}_{(1)}-\frac{\gamma_0\eta_0}{\sqrt{2}}\left\{
{\tilde x}+\frac{\sqrt{2}}{H}\left(\sigma^3+\Sigma\xi_{\rm (g)}^{1/2}\sin\chi_0\right)\right.\nonumber\\
&&\left.
-(x_{\rm (g)}-x_0)\left[
\frac{1}{\eta_{\rm (g)}}\tilde\nu^1+\frac{1}{\xi_{\rm (g)}^{1/2}\cos\chi_0}\tilde\nu^2
-\frac{H'}{H}\frac{\eta_{\rm (g)}}{\xi_{\rm (g)}^{1/2}\cos\chi_0}\left(\sigma^3+\Sigma\xi_{\rm (g)}^{1/2}\sin\chi_0\right)
\right]\right\}\,,\nonumber\\
{\mathcal C}_{(6)}&=&y_0{\mathcal C}_{(1)}-\frac{\gamma_0\eta_0}{\sqrt{2}}\left\{
{\tilde y}-\frac{\sqrt{2}}{H}\left(\sigma^2+\Sigma\xi_{\rm (g)}^{1/2}\cos\chi_0\right)\right.\nonumber\\
&&\left.
-(y_{\rm (g)}-y_0)\left[
\frac{1}{\eta_{\rm (g)}}\tilde\nu^1+\frac{1}{\xi_{\rm (g)}^{1/2}\sin\chi_0}\tilde\nu^3
+\frac{H'}{H}\frac{\eta_{\rm (g)}}{\xi_{\rm (g)}^{1/2}\sin\chi_0}\left(\sigma^2+\Sigma\xi_{\rm (g)}^{1/2}\cos\chi_0\right)
\right]\right\}\,.
\end{eqnarray}
where ${\mathcal C}_{(A)}=C_{(A)}/m$.
Evaluation at $u=0$ leads to
\begin{eqnarray}
\label{conserved0}
{\mathcal C}_{(1)}&=&-\frac{\gamma_0\eta_0}{\sqrt{2}}\,,\qquad 
{\mathcal C}_{(2)}=\gamma_0\xi_0^{1/2}\cos\chi_0\,,\qquad 
{\mathcal C}_{(3)}=\gamma_0\xi_0^{1/2}\sin\chi_0\,,\nonumber\\
{\mathcal C}_{(4)}&=&\gamma_0\xi_0^{1/2}(x_0\sin\chi_0-y_0\cos\chi_0)+\sqrt{2}\gamma_0\eta_0[\sigma_0^1+(1-\eta_0)\Sigma_0]\,,\nonumber\\
{\mathcal C}_{(5)}&=&-\gamma_0\eta_0\left[\frac{x_0}{\sqrt{2}}+\sigma_0^3+\Sigma_0\xi_0^{1/2}\sin\chi_0\right]\,,\qquad 
{\mathcal C}_{(6)}=-\gamma_0\eta_0\left[\frac{y_0}{\sqrt{2}}-\sigma_0^2-\Sigma_0\xi_0^{1/2}\cos\chi_0\right]\,,
\end{eqnarray}

\end{widetext}
where the initial conditions for the first order quantities have been fixed as $\tilde x^\alpha(0)=0=\tilde\nu^a(0)$.
Such a choice implies that the 4-velocity $U$ is initially tangent to the reference geodesic $U_{\rm (g)}$.

The set of equations (\ref{conserved}) and (\ref{conserved0}) gives five algebraic relations involving the unknown quantities $\tilde\nu^a$, $\tilde x$ and $\tilde y$ (plus one compatibility condition coming from ${\mathcal C}_{(4)}$, which is identically satisfied).
The solutions for the deviation velocities as well as spin-induced deviations in the transverse plane are given by Eqs. (\ref{soltildenu}) and (\ref{soltildex}), respectively.

\acknowledgments
DB thanks the INFN Section of Naples for partial support.


\begin{thebibliography}{00}

\bibitem{grifbook}
J.B. Griffiths, 
{\it Colliding Plane Waves in General Relativity} (Oxford University Press, Oxford, 1991).

\bibitem{baldwin}
O. Baldwin and G.B. Jeffrey,
{Proc.\ R.\ Soc.\ A} {\bf 111}, 95 (1926).

\bibitem{aichelburg}
P.C. Aichelburg,
{Acta\ Phys.\ Austr.} {\bf 34}, 279 (1971).

\bibitem{balasin}
H. Balasin, 
Class. Quantum Grav. {\bf 14}, 455 (1997).

\bibitem{vanholten}
A. Balakin, J.W. van Holten, and R. Kerner,
Class. Quantum Grav. {\bf 17}, 5009 (2000).

\bibitem{grifpod}
J.B. Griffiths and J. Podolski,
{\it Exact Space-Times in Einsteins General Relativity}
(Cambridge University Press, Cambridge, 2009). 

\bibitem{bgscat}
D. Bini and A. Geralico,
Phys. Rev. D {\bf 85}, 044001 (2012).

\bibitem{kirk}
L. Ball and J.G. Kirk, 
Astropart. Phys. {\bf 12}, 335 (2000).

\bibitem{math37} 
M. Mathisson,
{Acta Phys.\ Polon.} {\bf 6}, 163 (1937).

\bibitem{papa51} 
A. Papapetrou, 
{Proc.\ R.\ Soc.\ A} {\bf 209}, 248 (1951).

\bibitem{tulc59} 
W. Tulczyjew, 
{Acta\ Phys.\ Polon.} {\bf 18}, 393 (1959).

\bibitem{dixon64} 
W.G. Dixon, 
{Nuovo Cimento} {\bf 34}, 317 (1964).

\bibitem{dixon69}
W.G. Dixon, 
{Proc.\ R.\ Soc.\ A} {\bf 314}, 499 (1970).

\bibitem{dixon70}
W.G. Dixon, 
{Proc.\ R.\ Soc.\ A} {\bf 319}, 509 (1970).

\bibitem{dixon73}
W.G. Dixon, 
{Gen.\ Relativ.\ Gravit.} {\bf 4}, 199  (1973).

\bibitem{dixon74}
W.G. Dixon,  
{Phil.\ Trans.\ R.\ Soc.\ A} {\bf 277}, 59 (1974).

\bibitem{szekeres}
P. Bell and P. Szekeres, 
{Gen.\ Relativ.\ Gravit.} {\bf 5} 275 (1974).

\bibitem{grif}
J.B. Griffiths, 
Phys. Lett. A {\bf 54}, 269 (1975).

\bibitem{book}
F. de Felice and D. Bini, 
\textit{Classical Measurements in Curved Space-times} (Cambridge University Press, Cambridge, 2010).

\bibitem{abrasteg}
M. Abramowitz and I.A. Stegun (eds),
\textit{Handbook of Mathematical Functions with Formulas, Graphs, and Mathematical Tables} (9th printing),
(Dover, New York, 1972). 

\bibitem{maple}
Maple is a trademark of Waterloo Maple Inc..
Website at http://www.maplesoft.com

\bibitem{mathematica}
Wolfram Research, Inc., Mathematica.
Website at http://www.wolfram.com/mathematica

\bibitem{phinney}
E.S. Phinney, 
Mon.\ Not.\ R.\ Astron.\ Soc. {\bf 198}, 1109 (1982).

\bibitem{padman}
T. Padmanabhan,
J. Astrophys. Astr. {\bf 18}, 87 (1997).

\bibitem{quadrupgw}
D. Bini, P. Fortini, A. Geralico, and A. Ortolan, 
{Phys. Lett. A} {\bf 372} 6221 (2008).

\bibitem{mtw}
C.W. Misner, K.S. Thorne, and J.A. Wheeler, 
{\it Gravitation} (Freeman, San Francisco, 1973).

\bibitem{suckewer}
S. Suckewer,
Nature Phys. {\bf 7}, 11 (2011).

\bibitem{ES}
H. Stephani, D. Kramer, M.A.H. MacCallum, C. Hoenselaers, and E. Herlt, 
{\it Exact Solutions of Einstein's Field Equations}, 2nd edn. (Cambridge University Press, Cambridge, 2003).



\end{thebibliography}
\end{document}